\documentclass[aps,rmp,amsmath,amssymb,graphicx,longbibliography]{revtex4-1}

\usepackage{bm}
\usepackage{graphics}

\usepackage{caption}
\usepackage{textcomp}

\usepackage{graphicx}
\usepackage{epstopdf} 

\usepackage{floatrow}
\usepackage{subfig,amssymb}

\usepackage{color}

\begin{document}

\title{Response Formulae for $n$-point Correlations in Statistical Mechanical Systems and Application to a Problem of Coarse Graining}

\author{Valerio Lucarini}
\affiliation{Department of Mathematics and Statistics, University of Reading, Reading, RG66AX, UK}
\affiliation{Centre for the Mathematics of the Planet Earth, University of Reading, Reading, RG66AX, UK}
\affiliation{CEN - Institute of Meteorology, University of Hamburg, Hamburg, 20144, Germany}
\author{Jeroen Wouters}
\affiliation{Department of Mathematics and Statistics, University of Reading, Reading, RG66AX, UK}
\affiliation{Centre for the Mathematics of the Planet Earth, University of Reading, Reading, RG66AX, UK}
\affiliation{School of Mathematics and Statistics, The University of Sydney, Sydney, Australia}

\date{\today{}}

\begin{abstract}
Predicting the response of a system to perturbations is a key challenge in mathematical and natural sciences. Under suitable conditions on the nature of the system, of the perturbation, and of the observables of interest, response theories allow to construct operators describing the smooth change of the invariant measure of the system of interest as a function of the small parameter controlling the intensity of the perturbation. In particular, response theories can be developed both for stochastic and chaotic deterministic dynamical systems, where in the latter case stricter conditions imposing some degree of structural stability are required. In this paper we extend previous findings and derive general response formulae describing how $n-$point correlations  are affected by perturbations to the vector flow. We also show how to compute the response of the spectral properties of the system to perturbations. We then apply our results to the seemingly unrelated problem of coarse graining in multiscale systems: we find explicit formulae describing the change in the terms describing parameterisation of the neglected degrees of freedom resulting from applying perturbations to the full system. All the terms envisioned by the Mori-Zwanzig theory - the deterministic, stochastic, and non-Markovian terms - are affected at first order in the perturbation. The obtained results provide a more comprehesive understanding of the response of statistical mechanical systems to perturbations and contribute to the goal of constructing accurate and robust parameterisations and are of potential relevance for fields like molecular dynamics, condensed matter, and geophysical fluid dynamics. We envision possible applications of our general results to the study of the response of climate variability to anthropogenic and natural forcing and to the study of the equivalence of thermostatted statistical mechanical systems. 
\end{abstract}


\maketitle

\tableofcontents{}

\section{Introduction}
\label{intro}
\subsection{Response Theories}
Understanding how a system responds to perturbations is a key challenge in mathematical and natural sciences and has long been the subject of extensive analysis through formal, experimental, and numerical investigations. A fundamental step in the direction of developing a comprehensive response theory can be found in the early work of  \citet{K57} (see also \citet{kubo_statistical_1988}), who studied the impact of imposing weak perturbations to a statistical mechanical system originally at the thermodynamic equilibrium as described by the canonical ensemble. While the proposed theory had been criticised  from an early stage - see the famous argument by \citet{vK71} as discussed in \citet{marconi2008} - it has been extremely successful in describing many physical phenomena \cite{lucarini2005,marconi2008}.  The Kubo response theory leads to response formulae that express the change in the expectation value of a given observable $\Psi$ of the system as a perturbative series.  The zeroth order term is the expectation
value of the observable $\Psi$ in the unperturbed system, while the first order term, corresponding to the linear response, is expressed in terms of an explicitly determined causal Green's function, which contains comprehensive information on the interplay between the background dynamics of the system and the applied perturbation. It is important to  note that the Green's function itself is constructed as an expectation value of an observable on the unperturbed measure, with the ensuing effect that the unperturbed system contains the information needed for estimating its response to general forcings. This provides the basis for the cornerstone of Kubo's response theory, the so-called fluctuation-dissipation theorem (FDT), which links forced and free fluctuation in the linear perturbative regime.  This structure extends to higher order terms with a simple generalization, see e.g. \citet{LC12}

A basic pitfall of Kubo's approach in terms of physical applicability is the impossibility of dealing with  perturbations resulting from non-conservative forces. In fact, Kubo's theory does not allow for a consistent treatment of the energy budget of the perturbed system: in general, the external field will inject or subtract energy, so that in order to reach a well-defined steady state it is necessary to add a thermostat \cite{G1997,Cohen98,R2000}. The natural question is then whether a specific choice of the thermostat alters the computed linear response. Fortunately, as shown in \cite{evans_statistical_2008}, in the thermodynamic limit of a system with infinite number of particles, the choice of the thermostat does not alter the predictions of linear response theory: the sensitivity of macroscopic observables does not depend on the details of the microscopic dynamics.  

What is also unsatisfactory about the Kubo response theory is that mathematical rigour has been
missing in establishing whether the many limits involved in constructing the
response formulae are well defined. Additionally, no provision is given for computing the response of nonequilibrium systems to perturbations.

 \citet{R97,ruelle_nonequilibrium_1998,ruelle_review_2009} showed that
it is possible to establish a rigorous response theory for Axiom A maps and flows,
which possess  invariant Sinai-Ruelle-Bowen (SRB) measures. In other terms,
Ruelle showed that in the case of Axiom A systems the invariant measure is differentiable with
respect to the parameters controlling small modifications to the flow of the
system, and provided explicit expressions for the linear and higher order contributions to the response.  

Axiom A systems are indeed far from being typical dynamical systems,
but, according to the \textit{chaotic hypothesis} of Gallavotti and Cohen \cite{gallavotti_dynamical_1995,G96}, they
can be taken as effective models for  chaotic dynamical systems with many
degrees of freedom. Specifically, this means that when looking at macroscopic observables in \textit{sufficiently} chaotic (to be intended in a qualitative sense) high-dimensional  systems, it is expected that it is extremely hard to distinguish their properties from those of an Axiom A system, including some degree of structural stability. Note that the chaotic hypothesis can be seen as the natural extension of the ergodic hypothesis, which is the fundamental heuristic step needed to apply results of equilibrium statistical mechanics to interpret and predict the properties of  real  systems at equilibrium. 
Linear response is therefore expected to hold in practice for very general dynamical systems, while the known counter-examples are currently limited to low-dimensional non-uniformly expanding maps \cite{BS08,gottwald_spurious_2016}.

Axiom A systems corresponding to equilibrium physical systems possess an invariant measure that is absolutely continuous with respect to the Lebesgue measure because the phase space does not contract nor expand, as the flow is nondivergent. Axiom A systems featuring - on the average - a contraction in the phase space provide excellent mathematical models for nonequilibrium systems \cite{G06}. In this case, the invariant measure lives on a set with a Hausdorff dimension lower than the number of degrees of freedom of the system and is singular with respect to the Lebesgue measure, as a result of the contraction taking place in the stable manifold \cite{eckmann85}. Despite the geometrical complexity associated to the attractors of nonequilibrium systems, the Ruelle response theory, somewhat surprisingly, ensures that  differentiability can be established also in this case.




In the case of an equilibrium system, the Ruelle response theory allows for deriving the FDT. 
In nonequilibrium systems, instead, there is no one-to-one correspondence between forced and free fluctuations, as already suggested by \citet{lorenz79}:  \citet{R97,ruelle_nonequilibrium_1998,ruelle_review_2009} provides a mathematical explanation of this property, while a physical interpretation is given in, \textit{e.g.}, \citet{L08,L09,Lucarini:2011_NPG}. The basic idea is that while the natural fluctuations are able to substitute for the effect of the components of the forcing along the unstable manifold of the system, the impact of the components of the forcing along the stable manifold have no counterpart in the unperturbed system. 

Interestingly, while on one side there have been positive examples of applications of the FDT in nonequilibrium systems, like the climate, it is clear that, for a given class of forcing, the quality of the obtained response operator depends substantially on the chosen observable  \cite{gritsun_climate_2007,gritsun2008b}. In a recent paper, \citet{GL2017} have provided examples in a system of geophysical relevance of various scenarios supporting or not the applicability of the FDT to reconstruct the response of the system to perturbations. They have clearly shown that, indeed, when the applied forcing has a relevant projection on the stable manifold of the unperturbed system, the forced variability can have little resemblance to the natural one. In particular, the forcing can in some cases excite resonances corresponding to special dynamical features that are virtually unexplored by the unperturbed system, so that one can observe so called \textit{climatic surprises}. 

The difficulties in constructing \textit{ab-initio} the response operator using Ruelle's formulae come from the extremely different behaviour of the contribution coming from the unstable and stable manifold \cite{AM07a}. The computation of the contributions coming from the  stable directions give neither numerical nor conceptual problems. When the unstable directions are considered, problems emerge from the fact that contributions to the response come from integrals over time of exponentially growing functions, resulting from the presence of sensitive dependence on initial conditions. The ill--posedness of this operation is at the core of the \citet{vK71} criticism mentioned above. On the other side, response operators, as described in the next section, are constructed by  integrating over the statistical ensemble of the (unpertubed) system. Such an operation - under suitable conditions - regularises the previous divergences and explains why linear response is indeed well-posed. Nonetheless, obtaining in practice a stable estimate of the response operators from a finite number of ensemble members and from finite numerical simulations is far from obvious. We note that algorithms based on adjoint methods seem to  partially ease these issues \cite{EHL04,W13}. 

Convincingly good results in terms of climate prediction performed using the linear response theory have instead been obtained through bypassing the problem of constructing the response operator and using, instead, the formal properties of the Green's function \cite{Lucarini:2011_NPG,LBHRPW14,RLL14,LRL16}. Tests in simple models have emphasized that also the nonlinear response theory is extremely solid and amenable to numerical verification \cite{L09}.  

Modern methods of spectral theory  have provided different and elegant proofs and further generalizations of Ruelle's results. The response theory can be developed by comparing the Perron-Frobenius transfer operator \cite{B00} of the unperturbed and of perturbed system, thus focussing on the evolution of distributions rather than of individual trajectories - see e.g. \citet{liverani2006,BS08,B14c}. This approach has allowed the extension of Ruelle's results to systems more general than the Axiom A case, by focusing on constructing suitable Banach space of anisotropic distributions. 
The practical applicability of transfer operator-based methods for studying the response in high dimensional systems is still not entirely clear, as a result of the \textit{curse of dimensionality}, even if some optimism comes from the overall positive results obtained when severely reduced order models are considered \cite{Tantet2015a,Tantet2015b}.
Additionally, ideas borrowed from the theory of the transfer operator have proved extremely useful for studying the behaviour of geophysical systems in the vicinity of critical transitions, where the response theory breaks apart, decorrelation times become very long, and the presence of Ruelle-Pollicott resonances lead to the appearance of rough dependence of the system properties on the perturbation parameter \cite{chekroun2014}. 
Recently, explicit formulae based on simple matrix algebra have been proposed for computing the response of a finite state Markov chain to perturbations, thus providing a model for studying finer and finer partitions of actual phase spaces \cite{L2016}.

A different way to approach the problem of constructing a response theory can be followed by taking the point of view of stochastic dynamics, as proposed initially by \citet{Hanggi1975,Hanggi1977}; see a recent review by \citet{BM13}.  Adding (suitably chosen, typically gaussian white) noise on top of the deterministic dynamics allows to deal with invariant measures that are absolutely continuous with respect to Lebesgue and for making sure that the decay of correlations in the system is fast. As a result, some of the mathematical issues discussed above are automatically sorted out and, in particular, the FDT holds in all cases. Thanks to the presence of noise  it is possible to set a general framework for linear response theory in much greater regularity, including the case of infinite dimensional systems; see \citet{HM10} for a mathematically accurate study of linear response for stochastic system, where many  subtleties are sorted out. One needs to note, though, that while the presence of noise smoothens the invariant measure of the system, the weaker the noise, the harder it is for a numerical model to appreciate  such smoothness given the finite length numerical simulations and the finite size of the ensemble of performed simulations. 

\subsection{Parameterisation of a Coarse Grained Model: Stochasticity and Memory Effects}

Adding stochastic forcings on top of the deterministic dynamics should be justified on physical grounds and not used just as an \textit{ad hoc} assumption. A convincing way to motivate the introduction of a random component to the dynamics comes from the need of taking into account the effect of \textit{microscopic}, unresolved scales; see a mathematically  rigorous and complete treatment in \citet{CLW15a,CLW15b}. Along the lines of the early results by Mori and Zwanzig \cite{zwanzig_memory_1961,mori_transport_1965},  \citet{CLW15a,CLW15b} also clearly show that the construction of reduced order models unavoidably leads also to introducing non-Markovian terns in the surrogate dynamics of the variables of interest. 

The problem of constructing accurate and robust parameterisations for degrees of freedom that are hard to simulate explicitly is a crucial problem in a variety of scientific fields, and most notably in condensed matter physics \cite{Bhalla2016}, molecular dynamics \cite{Shinoda2007,baron2007,Kmiecik2016}, and in geophysical fluid dynamics \cite{Fetal15,Berneretal2016}. 

The situation in the case of atmospheric, ocean, and climate models is particularly complex because there is no clear gap (in terms of temporal and spatial scales) in variability of the fluid motions \cite{Ghil1987,Peixoto1992,LBHRPW14}. As a result, first, the approximation of infinite time separation between resolved and unresolved scales is unsatisfactory, so that the standard homogenization theory \cite{PS08} cannot be safely applied in this case. As a result, on one side the stochastic terms in the parameterisation cannot be represented as white noise, and the presence of memory effects leads additionally to the need to incorporate, in principle, non-Markovian terms in the dynamics.

Additionally, given the available numerical resolution at hand, one  always faces the problem of dealing with the so-called \textit{grey zone}, a range of scales where physical processes are only partially resolved \cite{Gerard2007}. Further, the parameterisation depends on where one defines the cutoff between resolved and unresolved scales of motion (practically often determined by the computational facilities at hand or the required length or number of the model runs), so that a painstaking process of tuning is in principle necessary each time the resolution of the model needs to be changed.  As  a result, the quest for self-adaptive parameterisation has been recently emphasized in the literature, see e.g. \citet{Arakawa2011,Park2014,Sakra2016}. Self-adaptivity is crucial for the goal of constructing models able to perform seamless prediction, \textit{i.e.} to be used  for weather forecast, seasonal prediction, and climate modelling \cite{Palmer2008}.

As for the scope of this paper, it is relevant to note that one can use the Ruelle response theory to compute explicitly the effect of small scale, fast degrees on freedom on the macroscopic ones.  In this case, the perturbation one studies using the results by Ruelle is exactly the coupling between the dynamics occurring at the different scales. One discovers that it is possible to derive an explicit parameterisation providing a deterministic, a stochastic, and a non-Markovian contribution to the dynamics of the variables of interest, thus obtaining a perturbative yet self-consistent closure to the problem \cite{WL12,WL13,WL16}. The various terms are constructed in terms of specific response operators at first and second order. Some first promising examples of applications of the theory and investigation of the skills of the parameterization schemes have been recently presented in models of various degrees of complexity \cite{Woutersetal2016,VissioLucarini2016,Demaeyer2017}.

\subsection{This Paper}

In this paper we set ourselves in the context of (possibly high-dimensional) chaotic deterministic dynamical systems, assume the chaotic hypothesis and, consequently, the applicability of the Ruelle response theory. We expect, nonetheless, that our results should apply also in the case of stochastic dynamics, apart from obvious changes in the notation. This paper has a twofold purpose and addresses an interdisciplinary audience. 

We first take a rather general point of view and note that most of the theoretical results presented in the literature focus on assessing the response of the system to perturbations in terms of changes of the expectation values of suitably defined observables. or, equivalently, of the invariant measure. This statement applies to both more heuristic and more rigorous studies, and both to approaches based on the framework of deterministic or stochastic dynamics. The \textit{elephant in the room} is, in our view, the lack (at least up to the authors' knowledge) of general explicit formulae predicting how the time-lagged correlations of observables change as a result of perturbations to the dynamics. Therefore, in this paper we  provide explicit linear response formulae for $n-$point time correlations of observables. As discussed below, in the general case treated here the response formulae become more involved than in the usual case of observables and one derived  new terms that cannot be framed, even in the case of unperturbed systems possessing smooth invariant measure,  in terms of the FDT. The possibility of having formulae for studying the response of higher order moments is quite attractive because it paves the way to asking how the statistical properties of the fluctuations of the system change as a result of the applied perturbation. In the specific case of climate dynamics, which is an application of special interest for the authors, this amounts to being able to address the question of how the climate variability changes in response to climate forcing \cite{Ghil2015}. This is a major and indeed open problem in the climate literature.

We then discuss  a - seemingly unrelated -  problem of interdisciplinary relevance, which was, in fact, the original driver of the investigation presented in this paper. We look into the problem of constructing reduced order models for multiscale systems and take advantage of the fact that, as mentioned above, it can be framed as an indeed nontrivial exercise that can be studied using response theory. Finding an accurate and efficient way to perform coarse graining in multiscale systems amounts to constructing a parameterised dynamics for the variables of interest (usually the large scale, slow ones) and is key to supporting the development of practically usable numerical models. A much desired quality of a parameterisation is its adaptivity with respect to changes in the properties of the system. In  previous publications \cite{WL12,WL13,WL16} we have introduced a general method for constructing parameterisations whose  main advantage is its adaptivity to the parameters describing the coupling and/or the time scale separation between the slow and fast scale of motion, whose lack is, instead, a key drawback of many other methods, and especially of the empirical ones. A basic issue, both at practical and at theoretical level, is to assess the robustness of a parameterisation with respect to small changes in the dynamics of the system.  In this paper, using the general results mentioned above, we are able to construct a \textit{response theory for the reduced order, coarse grained model}, and derive explicit formulae for the change of the various terms composing the parameterisation. This has relevance for the goal of constructing parameterisations able to adjust to small changes in the dynamics of the \textit{full} system. Note that such perturbations can also be considered as a representation of the model error: in this case, our results address the problem of understanding how the model error translates in the formulation of the reduced order model.


Being the numerical implementation and analysis of the response based parametersation  a topic that is in full development, the current extension of the theory consists mostly of formal calculations, at this stage. Numerical analysis will be the subject of future investigations.

The paper is organised as follows. In Section \ref{extresponse} we show how the response formulae are changed when the observable we are considering is also a function of the small parameter controlling the intensity of the forcing. In Section \ref{correlations} we use the result of Sect. \ref{extresponse} to present the extension of the response theory for the case of $n-$point correlations. We show in detail the calculations needed to reach general formulae that include, as special case, the usual response formulae for observables. The results contained in Sect. \ref{extresponse} might be of interest for experts in dynamical systems and statistical mechanics. In Section \ref{coarsegraining} we recapitulate how to construct parameterisations allowing for performing consistently coarse graining on multiscale systems and we show how the theory developed in Sect. \ref{correlations} allows for finding explicit formulae for the corrections to the parameterisations due to a perturbation applied to the full system. The results contained in Sect. \ref{coarsegraining} might be additionally of interest for scientists interested in specific applications of coarse graining methods, such as those working on the development of parameterisations for describing the coarse grained dynamics of systems of interest for, e.g. molecular dynamics or geophysical fluid dynamics. In Section \ref{conclusions} we discuss our results and present our conclusive remarks.


\section{A Simple Extension of the Standard Response Theory}
\label{extresponse}
Let's consider a continuous time Axiom A dynamical system \cite{eckmann85,ruelle89} defined on a compact $n$-dimensional manifold $\mathcal{M}$ of the form
\begin{equation}
\dot{\vec{x}}=\vec{F}(\vec{x})\label{baseflow}
\end{equation}
possessing an invariant measure $\rho_0$. We frame our results below in the setting of deterministic dynamical systems but we stress that equivalent equations will hold for stochastic differential equations.

The expectation value of a general observable $\Phi_0(\vec{x})$ on such a measure can be written as $\int_\mathcal{M} \rho_0(\mathrm{d}\vec{x})\Phi_0(\vec{x})$. We can also write the expectation value in a more compact form as $\rho_0(\Phi)$ or as $\langle \rho_0,\Phi_0\rangle $, where we stress that the expectation value is the result of applying a linear functional (the measure $\rho_0$) to the measurable function $\Phi_0$. 

Let $\vec{x}(t,\vec{x}_{0})$ be the flow from an initial condition $\vec{x}_{0}$, i.e. $\vec{x}(0,\vec{x}_{0})=x_{0}$ and $\vec{x}(t,\vec{x}_{0})$ satisfies \eqref{baseflow}. Then the Koopman operator $\Pi_0$ is the composition of an observable with the flow: $(\Pi_0(t) \Phi)(\vec{x}_0) = \Phi(x(t,x_0)) $. Under suitable conditions, one can express the Koopman operator as $\Pi_0(t)=\exp(\mathcal{L}_{(0)}t)$, where $\mathcal{L}_{(0)}=\vec{F}\cdot \vec{\nabla}$ is such that  $\dot\Psi=\mathcal{L}_{(0)}\Psi$ for all differentiable functions $\Psi=\Psi(\vec{x})$.
The Perron-Frobenius-Ruelle operator is the adjoint of the Koopman operator $\Pi_0^\top(t)$ and defines the push-forward of an initial measure $\rho^*$ so that $\rho(t,\rho^*)=\Pi_0^\top(t)\rho^*$, defined as follows:
\begin{equation} 
\int_\mathcal{M} \rho^*(\mathrm{d}\vec{x}_{0})\Phi_0(\vec{x}(t,x_{0}))=\langle \rho^*,\Pi_0(t)\Phi_0\rangle =\langle \Pi_0^\top(t)\rho^*,\Phi_0 \rangle=\langle \rho(t,\rho^*),\Phi_0\rangle=\int_\mathcal{M} \Pi_0^\top(t)\rho^*(\mathrm{d}\vec{x}_{0})\Phi_0(\vec{x}_{0}).
\end{equation}
Note that  we have $\Pi_0^\top(t)=\exp({\mathcal{L}^\top_{(0)} t})$, with $\mathcal{L}^\top_{(0)}\rho^*=\vec{\nabla} \cdot (\vec{F} \rho^*)$. Additionally, by definition, we have $\Pi_0^\top(t)\rho_0=\rho_0$  and, correspondingly, $\mathcal{L}^\top_{(0)} \rho_0=0$.

Let's now consider a small $\epsilon-$perturbation to the vector flow of the form  
\begin{equation}
\dot{\vec{x}}=\vec{F}(\vec{x})+\epsilon\vec{G}(\vec{x})\label{pertubflow}
\end{equation}
so that the perturbed flow possesses an invariant measure $\rho_\epsilon$, and one can define the perturbed Liouville operator as $\mathcal{L}_\epsilon=\mathcal{L}_{(0)}+\epsilon \mathcal{L}_{(1)}$, where $\mathcal{L}_{(1)} =\vec{G}\cdot \vec{\nabla}$. We also define the perturbed evolution and Perron-Frobenius-Ruelle operators as $\Pi_\epsilon(t)=\exp(\mathcal{L}_\epsilon t)$ and $\Pi^\top_\epsilon(t)=\exp(\mathcal{L}^\top_\epsilon t)$, respectively.

It is of clear relevance to be able to say under which conditions  for small values of $\epsilon$  it is possible to expand  $\langle\rho_\epsilon,\Phi_0 \rangle_\epsilon$ as follows:
\begin{equation}
\langle\rho_\epsilon,\Phi_0 \rangle =\langle\rho_0,\Phi_0 \rangle+ \epsilon \frac{\mathrm{d}}{\mathrm{d}\epsilon}\langle\rho_\epsilon,\Phi_0 \rangle  |_{\epsilon=0}+h.o.t.\label{perturb1}
\end{equation}
where $h.o.t.$ indicates higher order terms, and to find an explicit expression for the key quantity $\frac{\mathrm{d}}{\mathrm{d}\epsilon}\langle\rho_\epsilon,\Phi_0 \rangle  |_{\epsilon=0}$, which controls the first order correction of the expectation value. The Ruelle response theory says that if the unperturbed dynamical system $\dot{\vec{x}}=\vec{F}(\vec{x})$ is Axiom A and we consider a $C^3$ observable $\Phi_0(\vec{x})$, one can write    
\begin{equation}
\frac{\mathrm{d}}{\mathrm{d}\epsilon}\langle\rho_\epsilon,\Phi_0 \rangle  |_{\epsilon=0}
=\int_0^\infty  \mathrm{d}\tau\langle{\rho_0, \mathcal{L}_{(1)}
\exp{(\mathcal{L}_{(0)}\tau})\Phi_0}\rangle\label{response1}
\end{equation}
so that one can alternatively write $\rho_\epsilon=\rho_0+ \epsilon\frac{\mathrm{d}}{\mathrm{d}\epsilon} \rho_\epsilon|_{\epsilon=0} +h.o.t.$ where 
\begin{equation}
\frac{\mathrm{d}}{\mathrm{d}\epsilon} \rho_\epsilon|_{\epsilon=0}=\int_0^\infty  \mathrm{d}\tau \Pi_0^\top(t) \mathcal{L}^\top_{(1)}  \rho_0\label{response2};
\end{equation}
we write in this case $ \frac{\mathrm{d}}{\mathrm{d}\epsilon}\langle\rho_\epsilon,\Phi_0 \rangle  |_{\epsilon=0}= \langle \frac{\mathrm{d}}{\mathrm{d}\epsilon}\rho_\epsilon|_{\epsilon=0},\Phi_0 \rangle $. 

Note that if $\mathcal{L}_{(1)}=\mathcal{L}_{(0)}$, so that the perturbation is just a linear change in the time variable $t\rightarrow t(1+\epsilon)$, we have that $\frac{\mathrm{d}}{\mathrm{d}\epsilon} \rho_\epsilon|_{\epsilon=0}=0$ because $\mathcal{L}^\top_{(1)}  \rho_0=\mathcal{L}^\top_{(0)}  \rho_0=0$, from the definition of $\rho_0$. Note that rescaling time does not affect the expectation value of any observable at all orders of perturbations.

It is easy to generalise the problem to the case where the observable is a $C^1$ function of $\epsilon$ so that one can write the following expansion for small values of $\epsilon$: $\Phi_\epsilon=\Phi_0+ \epsilon\frac{\mathrm{d}}{\mathrm{d}\epsilon}  \Phi_\epsilon|_{\epsilon=0} +h.o.t.$. In this case, we have that 
\begin{equation}
\langle\rho_\epsilon,\Phi_\epsilon \rangle =\langle\rho_0,\Phi_0 \rangle+ \epsilon \frac{\mathrm{d}}{\mathrm{d}\epsilon}\langle\rho_\epsilon,\Phi_\epsilon \rangle  |_{\epsilon=0}+h.o.t.
\end{equation}
where the linear sensitivity can be expressed as:
\begin{equation}
 \frac{\mathrm{d}}{\mathrm{d}\epsilon}\langle\rho_\epsilon,\Phi_\epsilon \rangle  |_{\epsilon=0}=\langle \frac{\mathrm{d}}{\mathrm{d}\epsilon}\rho_\epsilon|_{\epsilon=0},\Phi_0\rangle+\langle \rho_0,\frac{\mathrm{d}}{\mathrm{d}\epsilon}\Phi_\epsilon|_{\epsilon=0}\rangle.
 \end{equation}
 where the first term corresponds to the usual response theory, and comes from the change of the dynamics of the system, while second term comes from the change of the definition of the observable as a function of $\epsilon$. 
 
Let's take a first simple and relevant example to illustrate the meaningfulness of this result. We consider as observable the divergence of the flow $\Phi_\epsilon = \vec{\nabla} \cdot(\vec{F}+\epsilon \vec{G})$ in Eq. \ref{pertubflow}. The expectation value of this observable is equal to the sum of the Lyapunov exponents of the system and can be interpreted as the opposite of its entropy production  \cite{ruelle89,G14}. We have that 
\begin{equation}
\frac{\mathrm{d}}{\mathrm{d}\epsilon}\langle\rho_\epsilon,\Phi_\epsilon \rangle  |_{\epsilon=0} =\int_0^\infty  \langle\rho_0,\mathrm{d}\tau \mathcal{L}_{(1)} \Pi_0(\tau)(\vec{\nabla}\cdot \vec{F})\rangle+\langle \rho_0,\vec{\nabla} \cdot\vec{G}\rangle.\label{EP}
\end{equation}
If the expectation value on the unperturbed measure of the divergence of perturbation flow is zero (or \textit{a fortiori} if the perturbation flow is divergence-free), the second term vanishes. See Appendix A for a discussion on the physical interpretation of Eq. \ref{EP}.

\subsection{Derivation of Response Formulae for $n$-point Correlations}
\label{correlations}

\subsubsection{Two-point Correlations}
We now consider as observable the product of the value two observables $\Psi_a$ and $\Psi_b$ taken as different times, \textit{i.e.}, without loss of generality $c_{\Psi_a,\Psi_b}(t):=\Psi_a(\vec{x})\Psi_b(\vec{x}(t))$. The expectation value of $c_{\Psi_a,\Psi_b}(t)$, is  $C_{\Psi_a,\Psi_b}(t)=\langle \rho_0,c_{\Psi_a,\Psi_b}(t)\rangle $,  the $t-$lagged correlation between $\Psi_a$ and $\Psi_b$. The local quantity $c_{\Psi_a,\Psi_b}(t)$ measures the joint fluctuations of the two observables $\Psi_a$ and $\Psi_b$ at different times but along the same orbit.

We consider the perturbed flow given in Eq. \ref{pertubflow}. The product $\Psi_a(\vec{x})\Psi_b(\vec{x}(t))$ can be written as $\Psi_a(\vec{x})\Pi_\epsilon(t)\Psi_b(\vec{x})$, so that we must add a lower index $\epsilon$ to the expressions $c_{\Psi_a,\Psi_b,\epsilon}(t)$ and to  $C_{\Psi_a,\Psi_b,\epsilon}(t)$.

In order to obtain an expression for $\frac{\mathrm{d}}{\mathrm{d}\epsilon}c_{\Psi_a,\Psi_b,\epsilon}(t)|_{\epsilon=0}$, we need to expand the Koopman for small values of $\epsilon$. Using the Dyson formalism, we have:
\begin{align}
\Pi_\epsilon(t)&=\exp{(\mathcal{L}_{(0)}t+\epsilon \mathcal{L}_{(1)} t)}=\Pi_{(0)}(t)+\epsilon \int_0^t \mathrm{d} \tau_2 \Pi_{(0)}(t-\tau_2)\mathcal{L}_{(1)}\Pi_{(0)}(\tau_2)+h.o.t.,\label{dyson}
\end{align}
where $h.o.t.$ indicates terms featuring higher powers of the parameter $\epsilon$. Note that the term proportional to $\epsilon$ in the right hand side of the previous equation is instrumental for deriving the desired result. We then have that the  linear response of the $t-$lagged time correlation between the two observables $\Psi_a$ and $\Psi_b$ can be written as:
\begin{equation}
\frac{\mathrm{d}}{\mathrm{d}\epsilon}C_{\Psi_a,\Psi_b,\epsilon}(t)|_{\epsilon=0}= \frac{\mathrm{d}}{\mathrm{d}\epsilon}\langle\rho_\epsilon,c_{\Psi_a,\Psi_b,\epsilon}(t)\rangle|_{\epsilon=0}=\langle \frac{\mathrm{d}}{\mathrm{d}\epsilon}\rho_\epsilon|_{\epsilon=0},c_{\Psi_a,\Psi_b,0}(t)\rangle+\langle \rho_0,\Psi_a\frac{\mathrm{d}}{\mathrm{d}\epsilon} \Pi_\epsilon(t)\Psi_{b}|_{\epsilon=0}\rangle.\label{bothterms}
\end{equation}
The first term on the right hand side gives to the correction of the local (in phase space) fluctuations computed according to the unperturbed dynamics due to the fact that the perturbation flow modifies the trajectories, and corresponds to what one would obtain with a naive application of the response theory for studying the change in the correlations of the system. The second term corresponds to the expectation value on the unperturbed dynamics of the change in the evolution law due to the presence of the $\epsilon-$perturbation.

In particular, we can write the first term as:
\begin{align}
\langle \frac{\mathrm{d}}{\mathrm{d}\epsilon}\rho_\epsilon|_{\epsilon=0},c_{\Psi_a,\Psi_b,0}(t)\rangle&=\langle \rho_0,\int_0^\infty \mathrm{d}\tau_1\mathcal{L}_{(1)}\Pi_{(0)}(\tau_1)\Psi_a\Pi_{(0)}(t)\Psi_b\rangle \nonumber\\
&=\int_0^\infty \mathrm{d}\tau_1\int_\mathcal{M} \rho_0(\mathrm{d}\vec{x})\vec{G}(\vec{x})\cdot\vec{\nabla}_{\vec{x}} (\Psi_a(\vec{x}(\tau_1))\Psi_b(\vec{x}(t+\tau_1))).\label{corr1}
\end{align}
Comparing with \citet{CL14}, we observe that this expression resembles a second order response term for regular observables, but, thanks to the presence of a slightly simpler functional form, can be brought to a FDT-like form by applying the operator $\mathcal{L}^\top_{(1)}$ to the unperturbed invariant measure $\rho_0$:
\begin{align}
\langle \frac{\mathrm{d}}{\mathrm{d}\epsilon}\rho_\epsilon|_{\epsilon=0},c_{\Psi_a,\Psi_b,0}(t)\rangle&=\int_0^\infty \mathrm{d}\tau_1\langle \mathcal{L}^\top_{(1)}\rho_0,\Pi_{(0)}(\tau_1)\Psi_a\Pi_{(0)}(t)\Psi_b\rangle\label{corr1b},
\end{align}
where we have an integral over one time variable of a three-point correlation. 

Instead, the second term in Eq. \ref{bothterms} can be written as: 
\begin{align}
\langle \rho_0,\Psi_a\frac{\mathrm{d}}{\mathrm{d}\epsilon} \Pi_\epsilon(t)\Psi_{b}|_{\epsilon=0}\rangle&=\langle \rho_0,\Psi_a\int_0^t \mathrm{d} \tau_2 \Pi_{(0)}(t-\tau_2)\mathcal{L}_{(1)}\Pi_{(0)}(\tau_2)\Psi_b\rangle \nonumber\\
&=\int_0^t \mathrm{d}\tau_2\int_\mathcal{M} \rho_0(\mathrm{d}\vec{x})\Psi_a(\vec{x})\vec{G}(\vec{x}(t-\tau_2))\cdot\vec{\nabla}_{\vec{x}(t-\tau_2)} (\Psi_b(\vec{x}(t)))\label{corr2}.
\end{align}
Note that this term vanishes if $t=0$ because in this case the function $c_{\Psi_a,\Psi_b,\epsilon}(t=0)$ is not anymore a function of $\epsilon$, and the usual response theory formulae apply. Due to the presence of a different time ordering in the operators, we cannot reframe Eq. \ref{corr2} in a FDT-like form.

We also wish to note that if the system is mixing and has rapid decay of correlations, both  terms given in the right hand side of  Eqs. \ref{corr1}-\ref{corr2} will tend to zero for large values of $t$.

In order to have a simple consistency test of our results, let's also take the special case seen above where $\mathcal{L}_{(1)}=\mathcal{L}_{(0)}$, \textit{i.e.}, we rescale the time variable $t\rightarrow t(1+\epsilon)$. In this case, the first term given in Eq. (\ref{corr1}) vanishes, because $\mathcal{L}^\top_{(0)}\rho_0=0$. This corresponds to what discussed before when looking at the response theory for observables. 

Instead, the second term reads $t \int_\mathcal{M} \rho_0(\mathrm{d}\vec{x})\Psi_a(\vec{x})\vec{F}(\vec{x(t)})\cdot\vec{\nabla}_{\vec{x}(t)} (\Psi_b(\vec{x}(t)))$. The (trivial) fact that rescaling time leads to a change in the correlations functions can be immediately derived by observing that
\begin{align}
&\frac{\mathrm{d}}{\mathrm{d}\epsilon}\langle\rho_0,\Psi_a(\vec{x})\Psi_b(\vec{x}(t(1+\epsilon)))\rangle|_{\epsilon=0}=t \langle  \rho_0,\Psi_a(\vec{x})\dot{\vec{x}}(\vec{x}(t))\cdot\vec{\nabla}\Psi_b(\vec{x}(t))\rangle=t \langle \rho_0,\Psi_a(\vec{x})\vec{F}(\vec{x}(t))\cdot\vec{\nabla}\Psi_b(\vec{x}(t))\rangle.
\end{align}
just as obtained above. 

\subsubsection{The General Case of $n$-point Correlations}

We now consider the case of general correlation functions. Take
\begin{align}
  &c_{\Psi_0, \Psi_1, \ldots,\Psi_{n-1}} (s_1, s_2, \ldots, s_{n_1}) = \Psi_0
   (\vec{x}) \Psi_1 (\vec{x} (s_1)) \ldots \Psi_{n-1} (\vec{x} (s_1 + \ldots +
   s_{n-1}))
\end{align}
and define the $n$-point correlation function for the perturbed system as:
\begin{align}
   C_{\Psi_0, \Psi_1, \ldots, \Psi_{n-1} ; \epsilon} (s_1, s_2, \ldots, s_{n-1}) &= \langle \rho_{\epsilon} ; c_{\Psi_0, \Psi_1, \ldots, \Psi_{n-1}} (s_1, s_2, \ldots, s_{n-1})) \rangle\nonumber\\
  & = \langle \rho_{\epsilon} ; \Psi_0 (\vec{x}) \Pi_{\epsilon} (s_1)
  \Psi_1 (\vec{x}) \Pi_{\epsilon} (s_2) \Psi_2 (\vec{x}) \ldots
  \Pi_{\epsilon} (s_{n-1}) \Psi_{n-1} (\vec{x}) \rangle \; .
\end{align}
We can then construct the following first order expansion for the $n$-point correlation as follows:
\begin{align*}
&C_{\Psi_0, \Psi_1, \ldots, \Psi_{n-1} ; \epsilon} (s_1, s_2, \ldots, s_{n-1}) = C_{\Psi_0, \Psi_1, \ldots, \Psi_{n-1} ; 0} (s_1, s_2, \ldots, s_{n-1}) + \epsilon\frac{\mathrm{d}}{\mathrm{d}\epsilon}C_{\Psi_0, \Psi_1, \ldots, \Psi_{n-1} ; \epsilon} (s_1, s_2, \ldots, s_{n-1})|_{\epsilon=0}+h.o.t. \\
 \end{align*}
The term proportional to $\epsilon$ is given by the sum of $n$ terms, the first one resulting from the linear correction to the measure, which corresponds to what one would naively obtain by applying the standard response theory, and the other $n-1$ terms resulting from the linear correction to each of the $n-1$ Koopman operators appearing in the definition of the $n$-point correlation function. We have: 

\begin{align}
   \frac{\mathrm{d}}{\mathrm{d}\epsilon}C_{\Psi_0, \Psi_1, \ldots, \Psi_{n-1} ; \epsilon} &(s_1, s_2, \ldots, s_{n-1})|_{\epsilon=0}= \int_0^{\infty} \mathrm{d} \tau
  \langle \rho_0, \mathcal{L}_1 \Pi_0 (\tau) \Psi_0 (\vec{x}) 
  \ldots \Pi_0 (s_{n-1}) \Psi_{n-1}
  (\vec{x}) \rangle \nonumber \\ 
  & + \sum_{k=1}^{n-1}\int_0^{s_k} \mathrm{d} \tau  \langle
  \rho_0, \Psi_0 (\vec{x}) \ldots \Pi_0 (s_k - \tau) \mathcal{L}_1 \Pi_0 (\tau)\Psi_k (\vec{x}) \ldots \Pi_0 (s_{n-1})\Psi_{n-1} (\vec{x})  \rangle\label{corrn}
\end{align}
As seen in the case of two-point correlations, the first term can be brought to a FDT-like form by applying the operator $\mathcal{L}^\top_{(1)}$ to the unperturbed invariant measure $\rho_0$, while the other terms have a more convolute structure.

\subsubsection{Change in the Spectral Properties of the System}

We can use the results presented before to draw interesting conclusions on how the spectral properties of the system under investigation change as a result of the $\epsilon-$ perturbation. Under suitable conditions of integrability, we have that $\mathcal{F}[C_{\Psi,\Phi}(t)]=P(\Psi,\Phi)=\mathcal{F}[(\Psi)]^*\mathcal{F}[(\Phi)]$, where $\mathcal{F}[g]=\hat{g}$ is the Fourier transform of $g$ and $f^*$ is the complex conjugate of $f$. With $P(\Psi,\Phi)=P(\Phi,\Psi)^*$ we indicate the co-spectrum of the two functions $\Psi$ and $\Phi$ (note the effect of the time lag). In particular, we have that if $\Psi=\Phi$, $\mathcal{F}[C_{\Psi,\Psi}(t)]=|\mathcal{F}[(\Psi)]|^2=|\hat\Psi|^2=P(\Psi,\Psi)$, which corresponds to the Khinchin-Wiener theorem. Thanks to the linearity of the Fourier transform, we can then derive the following expression from Eq. \ref{bothterms}:
\begin{equation}
\frac{\mathrm{d}}{\mathrm{d}\epsilon}P_\epsilon(\Psi_a,\Psi_b)|_{\epsilon=0}=\mathcal{F}\left[\langle \frac{\mathrm{d}}{\mathrm{d}\epsilon}\rho_\epsilon|_{\epsilon=0},c_{\Psi_a,\Psi_b,0}(t)\rangle\right]+\mathcal{F}\left[\langle \rho_0,\Psi_a\frac{\mathrm{d}}{\mathrm{d}\epsilon} \Pi_\epsilon(t)\Psi_{b}|_{\epsilon=0}\rangle\right],
\label{bothtermsfourier}
\end{equation}
where we have added a lower index to the cross-spectrum $P$ in order to keep track of the presence of the $\epsilon$-perturbation to the dynamics. Equation \ref{bothtermsfourier} provides the answer to the quite relevant question of how the spectral properties of the system change as a result of the presence of perturbations. Note that the first term on the right hand-side of Eq. \ref{bothtermsfourier} can be interpreted as cross-spectrum of the same observables $\Psi_a$ and $\Psi_b$ where the time statistics is computed according to the measure $\mathrm{d}\rho_\epsilon / \mathrm{d}\epsilon|_{\epsilon=0}$ (instead of the original invariant measure $\rho_0$). A simple dynamical-statistical interpretation for the second term is harder to provide, as the time-dependent operator appearing between the two observables leads to computing correlations (with respect to the unperturbed invariant measure $\rho_0$) between points in the phase space having no obvious dynamical link. See also the previous discussion around Eqs. \ref{corr1}-\ref{corr1b}. 

Note also that the linear response of higher order spectral properties of the system to the $\epsilon-$ perturbation can be derived by applying the $n-1$ dimensional Fourier transform in Eq. \ref{corrn}. This shows that our results allow for a more comprehensive understanding of the response of the system to perturbations than usual response theory.

We note that in \citet{lucarini2012} the problem of looking at the change of the spectral properties of a system had been approached from a different angle, studying the effect of  stochastic perturbations applied on top of deterministic chaotic dynamics. The main result obtained there is that one can establish a simple algebraic link between the change of the power spectrum of an observable (corresponding to the specific choice $\Psi_a=\Psi_b$ in terms of what presented here) and the squared modulus of the susceptibility referred to the same observable. 

\section{Response Formulae for Reduced Order Models}
\label{coarsegraining}
We find a useful application of the results detailed above in the special case of constructing parameterisations for reduced order models, along the lines of \citet{WL12,WL13,WL16}. Let's first recapitulate the main results obtained there and we shall then see how to apply the extended response theory described above to derive some new results. The idea is to derive formulae able to describe how the parameterisation changes as a result of perturbations applied to the full system, or, in other terms, how applying a perturbation changes the properties of the Mori-Zwanzig projection operator.

\subsection{Constructing the Projected Evolution Equations for Coarse Grained Variables}
We consider a high-dimensional chaotic dynamical system $\dot{\vec{z}}=\vec{F}_z(\vec{z})$ where $\vec{z}$ belongs to a compact manifold $\mathcal{Z}$, and then rewrite the dynamics by separating $\vec{z}$ into two subsets of variables, with $\vec{z}=[\vec{x};\vec{y}]$. Such a separation typically comes from the fact that we are interested in studying the properties of the $\vec{x}$ variables only,  corresponding to the coarse grained quantities of interest. Typically, the number of $\vec{y}$ variables is much larger than the number of $\vec{x}$ variables, and one would like to have a time-scale separation (or spectral gap) between the two sets of variables. Without loss of generality one can write:
\begin{align}
\dot{\vec{x}}&=\vec{F}_x(\vec{x})+\delta \vec{\Psi}_x(\vec{x},\vec{y})\label{fullsystema}\\
\dot{\vec{y}}&=\vec{F}_y(\vec{y})+\delta \vec{\Psi}_y(\vec{x},\vec{y})\label{fullsystemb}
\end{align}
\label{coarsegraininga}
where we have separated the part of the vector field ($\vec{\Psi}$) coupling the $\vec{x}$ and the $\vec{y}$ variables from the part of the vector field ($\vec{F}$) that drives independently the two groups of variables. We have also introduced the bookkeeping parameter $\delta$, which measures the strength of the coupling between the $\vec{x}$ and $\vec{y}$ variables. We wish to derive a reduced model for the $\vec{x}$ variables able to reproduce accurately (in some sense to be defined later) its statistical properties resulting from the full dynamics given in Eqs. \ref{fullsystema}-\ref{fullsystemb}. The Mori-Zwanzig theory allows for a exact and powerful yet implicit solution to this problem, obtained by formally removing  the evolution of the $\vec{y}$ variables. As a result, one obtains that it is possible to write the projected dynamics of the $\vec{x}$ variables as follows: 
\begin{equation}
\dot{\vec{x}}=\vec{F}_x(\vec{x})+ \vec{M}_\delta(\{\vec{x}\})\label{redsystem}
\end{equation}
where $\vec{M}$ contains both Markovian and non-Markovian components and provides the so-called parameterisation of the effect of the $\vec{y}$ variables on the $\vec{x}$ variables. The vector field $\vec{M}$ contains information on the average effect of the coupling between the $\vec{x}$ and $\vec{y}$ variables, on the impact of the fluctuations of the $\vec{y}$ variables, and on the memory effects due to nonlinear cross-correlations between the two groups of variables. 

Unfortunately, the explicit form of $\vec{M}$ is not in general available. In the limit of infinite time scale separation between the $\vec{x}$ and $\vec{y}$ variables, such that the $\vec{y}$ variables fluctuate infinitely faster than the $\vec{x}$ variables, it is instead possible to derive explicit results using the homogenization technique \cite{PS08}. 

One obtains that the $\vec{M}_\delta(\{\vec{x}\})$ term is given by the sum of a deterministic term, corresponding to the intuitive mean field effect, plus a white noise stochastic term, which describes the effect of the fluctuations, while the memory term disappears. Following  \citet{PS08}, one has that in physical systems the white noise should be interpreted in the sense of Stratonovich, as it should be considered as limiting case of a red noise having vanishing decorrelation time. 

This approach is extremely powerful and physically appropriate in all the situations where a substantial time-scale separation can be found between the two sets of variables. In situations, like in the case of climate dynamics, where there is no real spectral gap, the assumption of infinite time scale separation is risky.

In \citet{WL12,WL13,WL16} we have shown that, assuming that that $\delta$ is small (weak coupling hypothesis), it is possible to find an explicit expression of the Mori-Zwanzig corrections to the dynamics by performing a formal expansion of the Koopman operator in powers of $\delta$ and retaining the first two orders. The idea is to treat the coupling as a perturbation to the otherwise uncoupled dynamics of the $\vec{x}$ and $\vec{y}$ variables. One obtains that the surrogate dynamics of the $\vec{x}$ variables can be written as follows: 
\begin{equation}
\dot{\vec{x}}=\vec{F}_x(\vec{x})+ \delta\vec{M}_1(\vec{x})+ \delta\vec{M}_2(\{\vec{x}\})+\delta^2\vec{M}_3(\{\vec{x}\})\label{redsystem2}
\end{equation}
where $\vec{M}_1(\vec{x})$ is a determistic vector field, $\vec{M}_2(\{\vec{x}\})$ is a stochastic term constructed from the statistics of the fluctuations of the $\vec{y}$ variables, and $\vec{M}_3(\{\vec{x}\})$ is a non-Markovian term describing the fact that in the fully coupled  dynamics the current state of the $\vec{y}$ variables contains information on the  state of the  $\vec{x}$ variables at previous times. This result is in agreement with the general theory on model reduction proposed by \citet{CLW15a,CLW15b}. 

The explicit expressions for the terms on the right hand side of Eq. \ref{redsystem2} are obtained as follows. We start by defining $\rho_{u,y}$ as the invariant measure of the dynamical system $\dot{\vec{y}}=\vec{F}_y(\vec{y})$, where $u$ in the lower index refers to the fact the dynamics of $\vec{y}$ is uncoupled from the dynamics of $\vec{x}$, so that  $\langle  \rho_{u,y},\xi(\vec{y})\rangle$ the expectation value of a $\rho_{u,y}-$measurable observable $\xi(\vec{y})$. 

We then take the simplifying assumption that $\vec{\Psi}_x(\vec{x},\vec{y})=\vec{\Psi}^1_x(\vec{x})\vec{\Psi}^2_x(\vec{y})$ and $\vec{\Psi}_y(\vec{x},\vec{y})=\vec{\Psi}^1_y(\vec{x})\vec{\Psi}^2_y(\vec{y})$. As discussed in \citet{WL13,WL16}, such an assumption leads to simpler and easier to interpret formulae; yet, it does not really lead to a loss of generality, if one takes into account the possibility of expanding a function of both $\vec{x}$ and $\vec{y}$ variables as a sum of products of functions of separately $\vec{x}$ and $\vec{y}$ variables only, using a Schauder decomposition \cite{LT96}. 

The deterministic mean field term is given by:
\begin{align}
\vec{M}_1(\vec{x})&=\vec{\Psi}^1_x(\vec{x})\langle \rho_{u,y},\vec{\Psi}^2_x(\vec{y})\rangle\label{determi}
\end{align}
We introduce now the anomalies $\vec{\Psi'}^j_q(\vec{q})=\vec{\Psi}^j_q(\vec{q})-\langle \rho_{u,q},\vec{\Psi}^j_q(\vec{q})\rangle$ for $j=1,2$ and $q=x,y$. We have that the second term of the parameterisation can be written as:
\begin{align}
\vec{M}_2(\{\vec{x}\})&=\vec{\Psi}^1_x(\vec{x})\vec{\eta}\label{stochastic}
\end{align}
where $\vec{\eta}$ is a centered random process with time correlation given by 
\begin{equation}
C(\vec{\eta}(0),\vec{\eta}(t))=\langle \rho_{u,y},\vec{\Psi'}^2_x(\vec{y})\Pi_{0,x}(t)\vec{\Psi'}^2_x(\vec{y})\rangle,\label{correla}
\end{equation}
where $\Pi_{0,q}(t)$ indicates the Koopman operator of the $q=x,y$ variables in the uncoupled case with $\delta=0$, such $\Pi_{0,q}(t)A(\vec{q})=A(\vec{q}(t))$ for any function of the phase space $A$. Note that  the random process $\vec{\eta}$ is not unique, as, at the desired level of precision in terms of $\delta$, we only require that the noise is centered and with the above mentioned correlation properties.
Finally, the third term in the parameterisation provides the non-Markovian contribution to the reduced model and is given by
\begin{align}
\vec{M}_3(\{\vec{x}\})&=\int_0^\infty \mathrm{d}\tau \vec{h}(t-\tau,\Pi_0(t-\tau)\vec{x})\label{memory}
\end{align}
where the integration kernel $\vec{h}$ is written as
\begin{align}
\vec{h}(\sigma,\Pi_0(\sigma)\vec{x})&=\vec{\Psi}^1_y(\vec{x})\Pi_{0,x}(\sigma)\vec{\Psi}^1_x(\vec{x})\langle \rho_{u,y},\vec{\Psi}^1_y(\vec{y})\vec{\nabla}_{\vec{y}}\Pi_{0,y}(\sigma)\vec{\Psi}^2_x(\vec{y})\rangle.\label{memoryh}
\end{align}
A thorough interpretation of the three terms is reported in \citet{WL12,WL13,WL16}. 

We note that, using the Ruelle response theory, one also proves that up to second order in $\delta$ the invariant measure of the  dynamical system given in Eq. \ref{redsystem2} is the same as the $\vec{x}-$projection of the measure of the full dynamics given in Eqs. \ref{fullsystema}-\ref{fullsystemb}. Therefore, the parameterisation given in Eq. \ref{redsystem2} is effective in reproducing both the dynamical and the statistical properties of the full system.

Furthermore, as opposed to more common heuristic approaches, it performs - in the limit of small $\delta$ - consistently well no matter which observable $\Phi$ we are considering; it is, in this sense, universal and not targetted to a specific measure of skill. In \citet{Woutersetal2016,VissioLucarini2016,Demaeyer2017} the properties of parameterisations of models of different level of complexity obtained following this strategy are studied in detail. Note that in the limit of infinite time-scale separation between the  $\vec{x}$ and $\vec{y}$ variables, the homogeneization theory results are recovered and the non-Markovian term drops out.

\subsection{Impact of the Perturbations on the Parameterisation}
A basic problem often encountered when constructing parameterisations for unresolved processes is assessing the robustness of the reduced model with respect to small changes of the dynamics of the full system. When the dynamics of the full system is weakly perturbed with respect to reference conditions, one expects that also the reduced model undergoes small changes. In what follows, we define a set of response formulae able to predict how the various terms in Eqs. \ref{determi}-\ref{memory} defining the parameterisation change as a result of such a perturbation.  One needs to note that the presence of a small perturbation to the dynamics is usually interpreted as resulting from changes in the applied forcing applied or from changes in the value of some internal parameters. Alternatively, the small perturbation can be interpreted as caused by model error due to our incomplete knowledge of the system. We then consider the following system: 
\begin{align}
\dot{\vec{x}}&=\vec{F}_x(\vec{x})+\delta \vec{\Psi}_x(\vec{x},\vec{y})+\epsilon G_x(\vec{x})\\
\dot{\vec{y}}&=\vec{F}_y(\vec{y})+\delta \vec{\Psi}_y(\vec{x},\vec{y})+\epsilon G_y(\vec{y})\label{fullsystem2}
\end{align}
where we have included on the right hand side of the evolution equations a (small) perturbation vector field, whose intensity is controlled by the bookkeeping parameter $\epsilon$, while leaving the coupling unaltered with respect to the original system shown in Eqs. \ref{fullsystema}-\ref{fullsystemb}. In this case, the uncoupled model reads as
\begin{align}
\dot{\vec{x}}&=\vec{F}_x(\vec{x})+\epsilon G_x(\vec{x})\\
\dot{\vec{y}}&=\vec{F}_y(\vec{y})+\epsilon G_y(\vec{y})\label{uncsystem2}.
\end{align}
The reduced model, following Eq. \ref{redsystem2}, can be written as:
\begin{equation}
\dot{\vec{x}}=\vec{F}_x(\vec{x})+\epsilon G_x(\vec{x}) +\delta\vec{M}_{1,\epsilon}(\vec{x})+ \delta\vec{M}_{2,\epsilon}(\{\vec{x}\})+\delta^2\vec{M}_{3,\epsilon}(\{\vec{x}\})\label{redsystem3}
\end{equation}
where the dependence on $\epsilon$ is implicit for all terms except the trivial one. We now wish to expand the terms $\vec{M}_{1,\epsilon}(\vec{x})$, $\vec{M}_{2,\epsilon}\{\vec{x}\}$, and $\vec{M}_{3,\epsilon}\{\vec{x}\}$ in powers of $\epsilon$ and retain the 0$^{th}$ and 1$^{st}$ terms. This will lead us to the response formulae for the reduced order model. In order to do so, we define $\rho_{u,\epsilon,y}$ the invariant measure of the dynamical system in Eq. \ref{uncsystem2}, so that clearly $\rho_{u,\epsilon=0,y}=\rho_{u,y}$, and take advantage of the results contained in Sect. \ref{extresponse} in order to compute the linear response of expectation values of observables and correlations to the perturbation proportional to $\epsilon$.
Let's first look at the deterministic term introduced in Eq. \ref{determi}. We use Eqs. \ref{perturb1}-\ref{response1} to derive:
\begin{align}
\delta \vec{M}_{1,\epsilon}(\vec{x})&=\delta\vec{\Psi}^1_x(\vec{x})\langle \rho_{u,\epsilon,y},\vec{\Psi}^2_x(\vec{y})\rangle\nonumber\\
&=\delta\vec{M}_{1,\epsilon=0}(\vec{x})\nonumber\\
&+\delta\epsilon\vec{\Psi}^1_x(\vec{x})\int_0^\infty  \mathrm{d}\tau\langle{\rho_{u,y}, \mathcal{L}_{(1),y}
\exp{(\mathcal{L}_{(0),y}\tau})\vec{\Psi}^2_x(\vec{y}}\rangle+h.o.t.\label{response2b}
\end{align}
where $\vec{M}_{1,\epsilon=0}(\vec{x})$ is given in Eq. \ref{determi}, $\mathcal{L}_{(0),y} =\vec{F}_y\cdot \vec{\nabla}$, and $\mathcal{L}_{(1),y} =\vec{G}_y\cdot \vec{\nabla}$. Note that the correction term is proportional to $\epsilon$ so that, when we insert it in Eq. \ref{redsystem3}, it brings a contribution proportional to the product of the two perturbation parameters $\delta$ and $\epsilon$.

When looking at the modifications of the stochastic term given in Eq. \ref{stochastic}, we have that the $\epsilon$-correction to the dynamics of the $\vec{y}$ variables leads to a change in the correlation properties of the random process $\eta_\epsilon$. We obtain that
\begin{align}
\delta\vec{M}_{2,\epsilon}(\{\vec{x}\})&=\delta\vec{\Psi}^1_x(\vec{x})\vec{\eta_\epsilon}\label{stochasticb}
\end{align}
where we have that:
\begin{align}
&C(\vec{\eta_\epsilon}(0),\vec{\eta_\epsilon}(t))=C(\vec{\eta}_{\epsilon=0}(0),\vec{\eta}_{\epsilon=0}(t))+\epsilon \frac{\mathrm{d}}{\mathrm{d}\epsilon}\langle \rho_{u,\epsilon,y},\vec{\Psi'}^2_x(\vec{y})\Pi_{\epsilon,x}(t)\vec{\Psi'}^2_x(\vec{y})\rangle|_{\epsilon=0}+h.o.t.
\end{align}
with $C(\vec{\eta}_{\epsilon=0}(0),\vec{\eta}_{\epsilon=0}(t))$ given in Eq. \ref{correla};  using Eq. \ref{corr1}-\ref{corr2} we have 
\begin{align}
\frac{\mathrm{d}}{\mathrm{d}\epsilon}\langle \rho_{u,\epsilon,y},\vec{\Psi'}^2_x(\vec{y})\Pi_{\epsilon,x}(t)\vec{\Psi'}^2_x(\vec{y})\rangle|_{\epsilon=0}&=\int_0^\infty \mathrm{d}\tau_1\langle \rho_{u,y},\mathcal{L}_{(1),y}\Pi_{0,y}(\tau_1)\vec{\Psi'}^2_x(\vec{y})\Pi_{0,y}(t)\vec{\Psi'}^2_x(\vec{y})\rangle\nonumber\\
&+\int_0^t \mathrm{d} \tau_2\langle \rho_{u,y},\vec{\Psi'}^2_x(\vec{y})\Pi_{0,y}(t-\tau_2)\mathcal{L}_{(1),y}\Pi_{0,y}(\tau_2)\vec{\Psi'}^2_x(\vec{y})\rangle \label{parfluct}
\end{align}
The previous formula shows that the changes in the correlation of the noise due to the $\epsilon$-perturbation of the dynamics are non-trivial. In the limit of infinite time separation between the $\vec{x}$ and the $\vec{y}$ variables, such that the noise correlation is proportional to a Dirac's delta in both the unperturbed and perturbed system, the correction above results into a change of the constant in front of the $\delta$ by a factor proportional to $\epsilon$.

Finally, in order to construct the response formula for the term responsible for the non-Markovian part of the parameterisation, we need to evaluate the first order $\epsilon-$correction to the memory kernel $\vec{h}_\epsilon$, where 
\begin{equation}
\vec{M}_{3,\epsilon}\{\vec{x}\}=\int_0^\infty \mathrm{d}\tau \vec{h}_\epsilon(t-\tau,\Pi_\epsilon(t-\tau)\vec{x})\label{kernelpert}.
\end{equation}
By definition we have: 
\begin{align}
\vec{h}_\epsilon(\sigma,\Pi_\epsilon(\sigma)\vec{x})&=\vec{\Psi}^1_y(\vec{x})\Pi_{\epsilon,x}(\sigma)\vec{\Psi}^1_x(\vec{x})\langle \rho_{u,\epsilon,y},\vec{\Psi}^1_y(\vec{y})\vec{\nabla}_{\vec{y}}\Pi_{\epsilon,y}(\sigma)\vec{\Psi}^2_x(\vec{y})\rangle.\label{memoryh2}
\end{align}
and we wish to construct the following expansion:
\begin{align}
\vec{h}_\epsilon(\sigma,\Pi_\epsilon(\sigma)\vec{x})&=\vec{h}_{\epsilon=0}(\sigma,\Pi_0(\sigma)\vec{x})+\epsilon\frac{\mathrm{d}}{\mathrm{d}\epsilon}\vec{h}_\epsilon(\sigma,\Pi_\epsilon(\sigma)\vec{x})|_{\epsilon=0}+h.o.t.\label{propepsilon}
\end{align}
where $\vec{h}_{\epsilon=0}(\sigma,\Pi_0(\sigma)\vec{x})$ is given in Eq. \ref{memoryh}. On the r.h.s. of Eq. \ref{memoryh2} the parameter $\epsilon$ appears, reading from left to right, in the Koopman operator of the $\vec{x}$ variables, in the definition of the invariant measure, and in the Koopman operator of the $\vec{y}$ variables, thus implying that the term proportional $\epsilon$ in Eq. \ref{propepsilon} includes the sum of three separate corresponding contributions. 
The three terms are reported below in Eqs. \ref{parmemory1a}, \ref{parmemory1b}, and \ref{parmemory1c}, respectively: 
\begin{align}
\frac{\mathrm{d}}{\mathrm{d}\epsilon}\vec{h}_\epsilon(\sigma,\Pi_\epsilon(\sigma)\vec{x})|_{\epsilon=0}&=\vec{\Psi}^1_y(\vec{x})\int_0^\sigma \mathrm{d} \tau_0 \Pi_{0,x}(\sigma-\tau_0)\mathcal{L}_{(1),x}\Pi_{0,x}(\tau_0)\vec{\Psi}^1_x(\vec{x})\langle \rho_{u,y},\vec{\Psi}^1_y(\vec{y})\vec{\nabla}_{\vec{y}}\Pi_{0,y}(\sigma)\vec{\Psi}^2_x(\vec{y})\rangle\label{parmemory1a}\\
+&\vec{\Psi}^1_y(\vec{x})\Pi_{0,x}(\sigma)\vec{\Psi}^1_x(\vec{x})\int_0^\infty \mathrm{d}\tau_1\langle \rho_{u,y},\mathcal{L}_{(1),y}\Pi_{0,y}(\tau_1)\vec{\Psi}^1_y(\vec{y})\Pi_{0,y}(t)\vec{\Psi}^2_x(\vec{y})\rangle\label{parmemory1b}\\
+&\vec{\Psi}^1_y(\vec{x})\Pi_{0,x}(\sigma)\vec{\Psi}^1_x(\vec{x})+\int_0^\sigma \mathrm{d} \tau_2\langle \rho_{u,y},\vec{\Psi}^1_y(\vec{y})\Pi_{0,y}(\sigma-\tau_2)\mathcal{L}_{(1),y}\Pi_{0,y}(\tau_2)\vec{\Psi}^2_x(\vec{y})\rangle \label{parmemory1c}
\end{align}
It is interesting to note that the first contribution above in Eq. \ref{parmemory1a} is the only one involving the perturbation to the Liouville operator for the $x-$variables $\mathcal{L}_{(1),x}$. Correspondingly, it leads to a memory term in the definition of the kernel, which makes the overall non-Markovian term of the parameterisation more cumbersome; compare with Eq. \ref{kernelpert}.

The results presented here, albeit admittedly convoluted, show how it is in principle possible to construct the response theory for a reduced order model resulting from the coarse graining of higher dimensional system. In other terms, we find how one can construct a flexible parameterisation that can be explicitly adapted when the background system is altered, as a result of perturbations to the dynamics or taking into account the model error.

\section{Summary and Conclusions}
\label{conclusions}
Response formulae are extremely useful tools for predicting how the properties of statistical mechanical systems change as a result of perturbations. In practice, such perturbation can result from changes in the forcing applied to the system or to the internal parameters.  Mathematically solid response theories can be constructed both taking the point of view of chaotic deterministic dynamical systems - see \textit{e.g.} \cite{ruelle_review_2009,liverani2006} - and of stochastic dynamical systems - see \textit{e.g.} \cite{HM10}. The deterministic point of view faces the difficulty of requiring relatively stringent conditions of the nature of the flow, while the stochastic point of view permits deriving the desired results under more general conditions. The unavoidable price we pay in this latter case is that we should be able to justify the nature of the noise we use in our mathematical construction. For any practical use, the deterministic and the stochastic formulation of the problem are virtually equivalent.


In this paper we have extended the usual results of linear response theory by computing how the $n$-point correlations at different times of general smooth observables of the system under investigation change as a result of adding a weak perturbation to the vector flow. The obtained response formulae entail exactly $n$ different terms. The first term results results from the change in the invariant measure of the system, and is what one would guess from a naive use of response theory. The additional $n-1$ terms result from the linear correction to the Koopman operator of the system evaluated at all the $n-1$ consecutive intervals defining the  ordering of the time variables  in the argument of the correlation function. Such terms cannot be framed in any form similar to the FTD, as opposed to the first term. By taking advantage of the linearity of the Fourier transform, we are able to derive expressions describing how the spectral properties of the system are altered as a result of the presence of the perturbation. Formulae for second or higher order response to perturbations can also be obtained but are not presented here as they are rather complicated and do not add much for the scopes of this paper. 

We have then applied the general findings above to a problem of specific interest in the theory of coarse graining of multi-scale dynamical systems. From a truncation of the Mori-Zwanzig projection operator we can derive a parameterisation of the neglected degrees of freedom such that the resulting invariant measure of the surrogate system is identical to the projected measure of the full system up to second order in the parameter controlling the intensity of the coupling between the degrees of freedom of interest and the ones we want to neglect \cite{WL12,WL13,WL16}. One obtains that the parameterisation contains a deterministic component, a stochastic component, and a non-Markovian component, in agreement with the general theory of \citet{CLW15a,CLW15b}, and derives explicit expressions for the three terms. In this paper we have derived explicit expressions describing how the parameterisation changes as a result of a perturbation applied to the full system, or, in other terms, we have computed how the additional forcing projects in the reduced order model. Alternatively, one can see our results as a way to predict how the model error in the full system is translated as  error in the reduced order model.

One has to note that all the terms in \eqref{response2b}-\eqref{parmemory1c} are expectation values w.r.t. $\rho_{u,y}$, the uncoupled $y$ measure. Therefore, if we have access to such statistics, it is possible not only to construct a reduced model, but also to adapt it to account for small perturbations. Therefore, our results provide a basis for constructing general parameterisations for reduced order models that can be modified in order to account for changes in the dynamics of the full system. We suggest that this might be of relevance for fields such as condensed matter, molecular dynamics, and geophysical fluid dynamics, where the construction of accurate, flexible, and adaptive coarse graining procedures is of the uttermost relevance and urgency. In particular, in the case of geophysical fluid dynamics, our results might be useful for the construction of robust scale aware parameterisations, i.e, parameterisations that can be automatically or easily adapted to a changing grid resolution of the numerical model, which determines which physical processes can be explicitly resolved.

We will delve into the problem of implementing these results in specific numerical models and testing their accuracy in future investigations.

The formulae presented provide an overarching framework for understanding how higher order statistical moments of the systems are impacted by changes in the dynamics, and appear to be of general interest. In previous papers we showed that the Ruelle response theory is a tool of practical utility for approaching the problem of predicting climate change \cite{RLL14,LRL16}. Among the many possible applications of the results presented in this paper, we would to emphasise that the generalised response formulae introduced here allow for framing the question of how the climate variability responds to anthropogenic and natural forcings. This is a major and indeed open problem in the climate literature \cite{Ghil2015} and we will try to approach it in future studies.

An application of possible interest in the area of statistical mechanics deals with the study of the equivalence of perturbed Hamiltonian systems that are allowed to reach a steady state thanks to the coupling with thermostats described by different microscopic dynamics. In Appendix A we have briefly described the motivations behind the introduction of thermostats in physical systems. The formulae presented here  allow for computing explicitly the linear response of the correlations of the macroscopic physical variables of differently thermostatted perturbed Hamiltonian systems and then for checking whether an equivalence in the thermodynamic limit of such corrections exists, and if, so, how fast in terms of $N$, thus extending the results of \citet{evans_statistical_2008} for the case of linear response for physical observables.

\section*{Acknowledgments}
The results contained in this paper have been drafted during the Workshop \textit{Transport in Unsteady Flows:  from Deterministic Structures to Stochastic Models and Back Again} held on January 16-20 2017 at the Banff International Research Station, Banff, Canada. We thank the organizers and the institution for having made this possible. The authors wish to thank Judith Berner, Tamas Bodai, Mickael Chekroun,  Matteo Colangeli, Giovanni Gallavotti, Michael Ghil, Georg Gottwald, Cecile Penland, David Ruelle, Stephane Vannitsem, and Gabriele Vissio for many stimulating conversations on the topics discussed in this paper. VL wishes to thank the DFG TRR181 \textit{Energy Transfers in the Atmosphere and the Ocean} for partial financial support. The research leading to these results has received funding from the
European Community's Seventh Framework Programme (FP7/2007-2013) under
grant agreement No. PIOF-GA-2013-626210.

\section*{Appendix A: Thermostatted Systems}
A short note should be added in the case we are studying the response to perturbations of an $N$-particle system described by a Hamiltonian $H_0=K_0+V_0$ where $K_0=\sum_{j=1}^N \vec{p}_i^2/2m$ and $V_0=\sum_{j\neq i=1}^NV(|\vec{q}_i-\vec{q}_j|)$, where $\{\vec{q}_1,\ldots,\vec{q}_n,\vec{p}_1,\ldots,\vec{p}_n\}$ are the canonical variables, $m$ is the mass of the particles, and $V$ is the internal potential describing the interaction between the particles. The unperturbed system obeys the following equation of motions, for $i=1,\ldots,N$:
\begin{align}
\dot{\vec{q}}_i&=\vec{p}_i/m\nonumber\\
\dot{\vec{p}}_i&=-\vec{\nabla}_{\vec{q}_i}V_0.\label{hamil}
\end{align}
If we want to study the problem of deviations from equilibrium due to the application of an external (in general, non conservative) force $\epsilon\vec{B}(\vec{q}_i)$ acting on each particle, in order to keep physical well-posedness, we need to alter the vector flow as follows:
\begin{align}
\dot{\vec{q}}_i&=\vec{p}_i/m\nonumber\\
\dot{\vec{p}}_i&=-\vec{\nabla}_{\vec{q}_i}V_0+\epsilon\vec{B}(\vec{q}_i)-\alpha \vec{p}_i.\label{thermo}
\end{align}  
where $\alpha$ is a nontrivial friction coefficient describing the action of a thermostat \cite{G1997,Cohen98,R2000} that avoids the long-term accumulation or depletion of energy in the system and allows for the set up of a well-defined steady state. We consider here the case of deterministic thermostats.

As as example, choosing $\alpha= \epsilon \sum_{i=1}^N \vec{B}(\vec{q}_i)\cdot \vec{p}_i/\sum_{i=1}^N/(p_i^2/m)$, one obtains that the function $H_0$ is an invariant of the system given in Eq.  \ref{thermo}. Using such thermostatted equations of motions and considering as perturbation flow in Eq. \ref{pertubflow} $\epsilon\vec{G}(\vec{x})=(0,\ldots,0,\epsilon\vec{B}(\vec{q}_1)-\alpha \vec{p}_1,\ldots,\epsilon\vec{B}(\vec{q}_N)-\alpha \vec{p}_N)$\, where the perturbation affects only the evolution equations for the momentum variables, one recovers in Eq. \ref{EP} the correspondence between change in the phase space contraction rate and entropy production of the system mentioned above \cite{Cohen98}. Instead, neglecting the term responsible for the thermostatting, one instead derives from Eq. \ref{EP} the physically wrong result that the entropy production of an equilibrium system driven out of equilibrium by an external field vanishes.

Many functional forms can be given for $\alpha$, describing different ways of realising microscopically such long term balance. The equivalence of the thermostats means that in the thermodynamic limit $N\rightarrow\infty$ the expectation values of macroscopic physical observables does not depend on the choice of $\alpha$, with differences between the results obtained using different thermostats typically going to zero typically as  $1/N$ \cite{G1997,Cohen98,R2000,evans_statistical_2008,G14,GL14}. This property persists also when the sensitivity of the system is considered: in the thermodynamic limit the linear response of observables to perturbations is also independent of the choice of $\alpha$ \cite{evans_statistical_2008}. 



%


\end{document}